\documentstyle[12pt,aaspp4]{article}
\begin{document}
\def\msun{\hbox{${\cal{M}}_{\odot}$}}
\def\massA{\hbox{${\cal{M}}_A$}}
\def\massB{\hbox{${\cal{M}}_B$}}
\def\massAB{\hbox{${\cal{M}}_{A+B}$}}
%\slugcomment{to be submitted to {\it The Astronomical Journal},\hfill\break Version: 4/22/10}
\title{Speckle Interferometry at the U.S. Naval Observatory. XV.}

\author{Brian D. Mason, William I. Hartkopf and Gary L. Wycoff}
\affil{U.S. Naval Observatory \\ 3450 Massachusetts Avenue, NW, 
Washington, DC, 20392-5420 \\ Electronic mail: (bdm, wih, glw)@usno.navy.mil}

\begin{abstract}

%===========================================================================
Results of 2433 intensified CCD observations of double stars, made with the 
26-inch refractor of the U.S. Naval Observatory are presented. Each 
observation of a system represents a combination of over 2000 short-exposure
images. These observations are averaged into 1013 mean relative positions 
and range in separation from 0\farcs96 to 58\farcs05, with a mean separation
of 13\farcs50. This is the 15$^{th}$ in this series of papers and covers the
period 2008 January 3 through 2008 December 21. 

\end{abstract}


\section{Introduction}

%===========================================================================
This paper is the 15$^{th}$ in the series of papers from the U.S. Naval 
Observatory's speckle interferometry program, presenting results of 
observations obtained at the USNO 26 inch telescope in Washington, DC. Over
22,000 measures have now resulted from this program since its inception by 
Charles Worley, Geoff Douglass, and colleagues in the early 1990s (see 
Douglass et al.\ 1997).

%===========================================================================
From 2008 January 3 through 2008 December 21, the 26 inch telescope was 
used on 72 of 256 (28\%) scheduled nights. Based on this raw number, it 
might be concluded that climatic conditions have degraded over those 
summarized in Mason et al.\ (2006b; see especially their Figure 1). However,
there are currently fewer available observers, so nights judged as marginal 
are not covered to give preference to better nights. Consequently, while the
number of nights covered and total observations may have dropped, the 
observations per night metric has gone up. While most nights were lost due 
to weather conditions, time was also lost due to equipment upgrades and to 
personnel observing on other telescopes. Since our primary speckle camera 
was in use at other facilities during this period, all of these observations
were obtained with the secondary camera, described by Mason et al.\ (2007). 

%===========================================================================
Due to a larger plate scale and the lack of as many correcting optics as
contained in the primary camera, pairs observed with this camera are at much
wider separations. In fact, most of the systems observed with this camera 
have separations well beyond the regime in which there is any expectation of
isoplanicity, so for purposes of the $``$method" column of the WDS, we 
classify the observing technique for all of these measures as just ``CCD 
astrometry", rather than speckle interferometry. Despite this 
classification, there is an expectation that the resulting measurements have
smaller errors than would be expected for classical CCD astrometry. Each 
measurement is the result of many hundreds of correlations per frame, and up
to several thousand frames per observation. This ensemble of observations is
then processed and measured using the conventional directed vector 
autocorrelation techniques used by the CHARA and USNO speckle teams for over
20 years.

%===========================================================================
While individual nightly totals varied substantially (from 2 to 80 objects 
per night) these efforts yielded a total of 2433 observations and 2289 
resolutions (i.e., usable double star measurements). After removing marginal 
observations, calibration data and tests, a total of 1684 measurements 
remained, which were grouped into 1013 mean positions. Included in these are 
36 confirmations of binaries with only one previous observation. While some 
of these are relatively recent discoveries of the {\it Hipparcos} or {\it 
Tycho} missions (ESA 1997), some of these pairs had remained unconfirmed for
over 100 years. 

%===========================================================================
Observing list construction and calibration procedures remain the same as 
those described for the secondary camera in Mason et al.\ (2007). The 
plate scale of the secondary camera is not appropriate for the slit-mask 
calibration technique used for the primary camera, so observations of 
well-observed doubles were used instead. Evaluation of the ensemble of the 
tabulated $O-C$ in Table 2 allows the error to be grossly characterized as 
$\pm$1\fdg0 in position angle ($\theta$) and $\pm$1\% in separation 
($\rho$).

\section{Results}

%===========================================================================
Table 1 presents the mean relative position of the members of 754 systems 
having no published orbital elements. The first two columns identify the 
system by its epoch-2000 coordinates and discovery designation. Columns 
3-5 give the epoch of observation (expressed as a fractional Besselian 
year), position angle (in degrees), and separation (in arcsec). Note that 
the position angle has not been corrected for precession, and is thus based 
on the equinox for the epoch of observation. Objects whose measures are of 
lower quality are indicated by colons following the position angle and 
separation. These lower-quality observations may be due to one or more of 
the following factors: close separation, large $\Delta$m, one or both 
components very faint, a large zenith distance, and poor seeing or 
transparency. They are included primarily due to either the confirming 
nature of the observation or the number of years since the last measured 
position. The sixth column indicates the number of independent measurements 
(i.e., observations obtained on different nights) contained in the mean, and
the seventh column flags any notes. The 754 mean positions in Table 1 have 
an average separation of 12\farcs29. 

%===========================================================================
The most common note indicators are either ``C,'' indicating a confirming 
observation, or a number (N) indicating the number of years since the system
was last measured. This is only given for systems with N $\ge$ 50 years. 
Thirty-six systems are confirmed here. Since priority is given to both 
unconfirmed systems and to systems not observed recently, the time since 
last observation can be surprisingly large; for the systems in Table 1 the 
average time since the last observation is 12 years (67 years for those 
measures of reduced accuracy). Thirty-two systems had not been observed in 
50 years or more and 12 had not been observed for at least a century. The 
maximum such time span was 114 years for the pair SEI 374, which was first 
measured off {\it Carte du Ciel} plates by Scheiner (1908). 

%===========================================================================
Table 2 presents the mean relative positions for 254 binary star systems 
with published orbital determinations or linear solutions. The first six 
columns are identical to the corresponding columns of Table 1. Columns 7 and
8 give O$-$C residuals (in $\theta$ and $\rho$) to the determination 
referenced in Column 9. Like Table 1, the position angle has not been 
corrected for precession, however, the residual is relative to the precessed
value. The reference is to either a published orbit or a determination in 
the ``Catalog of Rectilinear Elements'' (Hartkopf et al.\ 2006), indicated 
by the letter {\bf L}. The objects in Table 2 tend to be more frequently 
observed than those in Table 1, with a mean separation of 17\farcs34, and a
mean time interval since last observation of only 1.4 yr. For those 50 pairs
with orbits, the mean separation is 4\farcs92 and time since last 
observation is 0.6 yr; both values are significantly less than 20\farcs39 
and 1.6 yr for those 204 pairs with linear solutions. This is appropriate; 
given measurements of equal quality, linear systems are more well suited for
calibration and should not require the same high observing cadence as orbit 
systems. As discussed in Mason et al.\ (2008, \S 3), double stars in which 
one component has high proper motion can be ascertained as either optical or
physical using a single high precision measure, assuming sufficient time has
passed since the last observation. Based on this methodology, none of the 
pairs in Table 2 are identified as optical but three are identified as 
physical (i.e., common proper motion). These are indicated in Table 1 with 
notes.

\section{Double Stars not Found}

%===========================================================================
Table 3 presents 12 systems which were observed but for which no companions 
were detected. Possible reasons for nondetection include orbital or 
differential proper motion making the binary too close or too wide to 
resolve at the epoch of observation, a larger than expected $\Delta$m, 
incorrect pointing, and misprints and/or errors in the original reporting 
paper. It is hoped that reporting these will encourage other double star 
astronomers to either provide corrections to the USNO observers or verify 
the lack of detection. Indeed, some of these systems in Table 3 may be 
optical doubles as described in the preceding section.

\acknowledgements

%===========================================================================
The continued instrument maintenance by the USNO instrument shop, Gary 
Wieder, John Evans, Tie Siemers and David Smith, makes the operation of  
this vintage telescope a true delight. Thanks also to Ted Rafferty (USNO,
retired) for his assistance with equipment upgrades and maintenance, and the
foresight to initiate the backup camera project. This research has made use 
of the SIMBAD database, operated at CDS, Strasbourg, France. Thanks are also
extended to Ken Johnston, Ralph Gaume and the U.\S.\ Naval Observatory for
their continued support of the Double Star Program.

%\documentstyle[aj_pt4]{article}
%\begin{document}

\scriptsize

%\end{document}

%\documentstyle[aj_pt4]{article}
%\begin{document}

\scriptsize
\begin{deluxetable}{ll@{~}rccccc}
\tablenum{3}
\tablewidth{0pt}
\tablecaption{Double Stars Not Found}
\tablehead{
\colhead{Coordinate} & 
\multicolumn{2}{c}{Discoverer} &
\multicolumn{3}{c}{Most Recent Published Observation} &
\multicolumn{2}{c}{Published Magnitude} \nl
\colhead{$\alpha,\delta$ (2000)} &
\multicolumn{2}{c}{Designation} & 
\colhead{Date} &
\colhead{Position Angle} &
\colhead{Separation} &
\colhead{Primary} &
\colhead{Secondary} \nl
\colhead{~} &
\multicolumn{2}{c}{~} & 
\colhead{~} &
\colhead{($\theta$)} & 
\colhead{($\rho$)} &
\colhead{~} &
\colhead{~} \nl
}
\startdata
15086$+$1103 & HJ  &  247 & 1820 & \phn50 &    15.0 &    10.0 & 11.0 \nl
15399$+$0227 & SLE &  926 & 1910 & \phn48 &    15.6 &    11.6 & 12.6 \nl
17053$+$1947 & SLE &    6 & 1982 &    138 &    20.3 & \phn9.5 & 11.0 \nl
17128$+$2433 & POU & 3264 & 1892 &    299 &    15.8 &    11.6 & 11.9 \nl
17184$+$0445 & SLE &   18 & 1982 &    333 &    17.7 &    11.0 & 11.5 \nl
17219$-$1846 & SLE &   14 & 1982 &    290 &    16.9 &    10.0 & 10.2 \nl
18444$+$2015 & ALD &    4 & 1916 &    339 & \phn3.9 &    10.9 & 11.0 \nl
18553$+$1217 & HEI &  564 & 1987 &    233 & \phn2.8 &    10.0 & 10.2 \nl
19258$+$4556 & HJ  & 1400 & 1909 &    181 & \phn5.8 &    10.4 & 11.1 \nl
20419$+$3143 & SEI & 1221 & 1896 &    254 & \phn2.0 &    10.9 & 11.1 \nl
21098$+$3143 & ES  & 2440 & 1930 &    202 & \phn5.0 &    10.9 & 11.9 \nl
22246$+$2337 & POU & 5698 & 1891 & \phn90 &    11.6 &    10.8 & 11.1 \nl
\enddata
\end{deluxetable}
%\end{document}

\end{document}